\documentclass[
fleqn]{elsart}
\usepackage{amsmath,amssymb,dcolumn}

\def\be{\begin{equation}}
\def\ee{\end{equation}}
\def\bea{\begin{eqnarray}}
\def\eea{\end{eqnarray}}
\def\ba{\begin{array}}
\def\ea{\end{array}}
\def\nn{\nonumber}

\begin{document}
\begin{frontmatter}

\title{Hawking radiation of Dirac particles via tunnelling from rotating black holes in de Sitter spaces}
\author[1]{De-You Chen},
\author[2]{Qing-Quan Jiang\corauthref{cor}}
\ead{jiangqq@iopp.ccnu.edu.cn}
\corauth[cor]{Corresponding author.}
and
\author[1]{Xiao-Tao Zu}
\address[1]{Department of Physics and Electron, University of Electric Science
and Technology of China, Chengdu 610054, Sichuan, China}
\address[2]{Institute of Particle Physics, Central China Normal University,
Wuhan, Hubei 430079, People's Republic of China}

\begin{abstract}
Hawking radiation from black hole horizon can be viewed as a quantum tunnelling process,
and fermions via tunnelling can successfully recover Hawking temperature.
In this paper, considering the tunnelling particles with spin $1/2$ (namely, Dirac particles),
we further improve Kerner and Man's fermion tunnelling method to
study Hawking radiation via tunnelling from rotating black holes
in de Sitter spaces, specifically including that from Kerr de Sitter black hole and Kerr-Newman de Sitter black hole.
As a result, Hawking temperatures at the event horizon (EH) and the cosmological horizon (CH)
are well described via Dirac particles tunnelling.
\end{abstract}

\end{frontmatter}
\newpage

\section{Introduction}

Since Stephen Hawking proved black holes could radiate
thermally\cite{r1}, people have attempt to provide several
different methods to correctly derive Hawking temperatures of
black holes. In these modes, a semi-classical tunnelling one,
first put forward by Kruas and Wilczek \cite{r2} and then
formulated by various researchers\cite{r3,r4,r5}, recently
attracts many people's attention \cite{r6}. Here the derivation of
Hawking temperature mainly depends on the computation of the
imaginary part of the action for the classically forbidden process
of s-wave emission across the horizon. Normally there are two
approaches to obtain the imaginary part of the action. One, first
used by Parikh and Wilczek\cite{r3} and later broadly discussed by
many papers\cite{r6}, is called as the Null Geodesic Method, where
the contribution to the imaginary part of the action only comes
from the integration of the radial momentum $p_r$ for the emitted
particles. The other method regards the action of the emitted
particles satisfies the relativistic Hamilton-Jacobi equation, and
solving it yields the imaginary part of the action\cite{r4}, which is an extension
of the complex path analysis proposed by Padmanabhan et al\cite{r5}.

Since the tunnelling method has been successfully applied to deal with Hawking radiation of
black holes, a lot of work shows its validity\cite{r6}. But most of them
are focus on studying Hawking radiation of scalar particles tunnelling from different-type black holes.
In fact, a black hole can radiate all types of particles, and Hawking radiation
contains contributions of both scalar particles and fermions with all spins.
Recently, Kerner and Mann have studied Hawking radiation of spin 1/2 fermions
for Rindler space-time and that for the uncharged spherically symmetric
black holes via quantum tunnelling method\cite{r7}. The result shows that Hawking
temperature can also be exactly obtained by fermions tunnelling from black hole horizon.
In this paper, we improve Kerner and Mann's method to deal with Hawking radiation of Dirac
particles via tunnelling from rotating black holes in de Sitter space, specifically
including that from Kerr de Sitter black hole and Kerr-Newman de Sitter black hole. The result
shows that Hawking temperatures at the EH and the CH can well be recovered Dirac particles via
tunnelling from Kerr de Sitter black hole and Kerr-Newman de Sitter black hole. Subsequently,
this method appears in many papers to discuss Hawking radiation from black holes\cite{M12,M9,M10,M11}.

Researches on black holes with a positive cosmological constant become important
due to the following reasons: (i)The recent observed accelerating expansion of
our universe indicates the cosmological constant might be a positive one \cite{r14};
(ii) Conjecture about de Sitter/conformal field theory (CFT) correspondence\cite{r15}.
For black holes in de Sitter spaces, particles can be created at both the event
horizon (EH) and the cosmological horizon (CH), where however exists different
tunnelling behaviors. At the EH, outgoing particles tunnel from black hole horizon to form
Hawking radiation, and incoming particles can fall into the horizon along
classically permitted trajectories. At the CH, outgoing particles can fall
classically out of the horizon, and incoming particles tunnel into the horizon to
form Hawking radiation.

The remainders of this paper are outlined as follows. In Sec.\ref{kds},
Hawking radiation of Dirac particles from Kerr de Sitter
black hole has been studied by improving Kerner and Man's method, and the temperatures at the
event horizon(EH) and that at the cosmological horizon(CH) are both well recovered.
Sec.\ref{knds} takes Kerr-Newman de Sitter black hole as an example to once
again check the validity of fermions tunnelling model. Sect.\ref{SSDB} ends up with
some discussions and conclusions.

\section{Dirac particles tunnelling from Kerr-de Sitter black hole} \label{kds}

In this section, we develop Kerner and Man's fermion tunnelling method to
study Hawking radiation of Dirac particles via tunnelling from rotating
Kerr-de Sitter black hole. The metric of Kerr black hole in de Sitter
spaces (Kerr de Sitter black hole) is given by \cite{r8}
\bea
\label{eq1}
ds^2 &=& - \frac{\Delta }{\rho ^2}\left[ {dt - \frac{a\sin ^2\theta
}{\Xi }d\varphi } \right]^2+ \rho
^2\left(\frac{dr^2}{\Delta } + \frac{d\theta ^2}{\Delta _\theta }
\right)  \nn\\
&+& \frac{\Delta _\theta \sin ^2\theta }{\rho ^2}\left[ {adt - \frac{r^2 + a^2}{\Xi }d\varphi } \right]^2.
\eea
where
\bea
\label{eq2}
&&\Delta = \left( {r^2 + a^2} \right)\left( {1 - r^2l^{ - 2}} \right) - 2Mr,  \nn\\
&&\Xi = 1 + a^2l^{ - 2}, \nn\\
&&\rho ^2 = r^2 + a^2\cos ^2\theta, \nn\\
&&\Delta _\theta = 1 + a^2l^{ - 2}\cos ^2\theta.
\eea
Here $a$ is rotational angular momentum parameter, and $l$ is a
constant related to the cosmological factor as $\Lambda = 3l^{ -2}$.
 There are four roots for $\Delta = 0$, and the largest
positive root $r_c$ corresponds to the cosmological horizon(CH) of
the black hole, the minimal positive root $r_i $ denotes the
Cauchy horizon of the black hole, the intermediate $r_h $ is the
event horizon(EH) of the black hole and $r_ - $ is the negative
root. For simplicity to choose the approximate metrics $\gamma^\mu$, we carry out
the dragging coordinate transformation as $\phi = \varphi - \Omega t$, where
\begin{equation}
\label{eq3}
\Omega = \frac{\left[ {\Delta _\theta \left( {r^2 + a^2} \right) -
\Delta } \right]a\Xi }{\Delta _\theta \left( {r^2 + a^2} \right)^2 -
\Delta a^2\sin ^2\theta },
\end{equation}
to the line element (\ref{eq1}), then the new form of the metric is
\begin{equation}
\label{eq4} ds^2 = - F\left( r \right)dt^2 + \frac{\mbox{1}}{G\left(
r \right)}dr^2 + K^2\left( r \right)d\theta ^2 +
H^2\left( r \right)d\phi ^2,
\end{equation}
where
\bea
\label{eq5}
&& F\left( r \right) =\frac{\Delta \Delta _\theta \rho
^2}{\Delta _\theta \left( {r^2 + a^2} \right)^2 - \Delta
a^2\sin ^2\theta }, \quad G\left( r \right) = \frac{\Delta }{\rho
^2},\nn\\
&& H^2\left( r \right) = \frac{\sin ^2\theta }{\Xi ^2\rho
^2}\left[ {\left( {r^2 + a^2} \right)^2\Delta _\theta -
\Delta a^2\sin ^2\theta } \right], \quad K^2\left( r \right) =
\frac{\rho ^2}{\Delta _\theta }.
\eea

The line element (\ref{eq4}) has many superior properties: (i)
the metric is regular at the EH and the CH; (ii) the EH and the CH
are respectively coincident with its infinite red-shift surface; (iii)
it satisfies Landau's condition of the coordinate clock synchronization.
All these features are very helpful to study Hawking radiation from the rotating
Kerr de Sitter black hole via quantum tunnelling method proposed by
Parikh and Wilczek\cite{r3}. In our case, this coordinate transformation would
provide us much more simplicity to choose the matrices $\gamma ^\mu $.
To view Hawking radiation of Dirac particles, we must first introduce its motion equation.
For Dirac particles in black hole space-time, its motion equation can be written as
\begin{equation}
\label{eq6} i\gamma ^\mu \left( {\partial _\mu + \Omega _\mu }
\right)\psi + \frac{m}{\hbar }\psi = 0,
\end{equation}
where $\Omega _\mu = \frac{i}{2}\Gamma _\mu ^{\alpha \beta } \sum
_{\alpha \beta } $ , $\sum _{\alpha \beta } = \frac{i}{4}\left[
{\gamma ^\alpha ,\gamma ^\beta } \right]$, and the matrices $\gamma ^\mu $
satisfy $\left\{ {\gamma ^\mu ,\gamma ^\nu } \right\} =
2g^{\mu \nu }\times I$.
There are many different ways to choose the $\gamma
^\mu $ matrices, and for simplicity our choice is
\bea
 \label{eq7}
&&\gamma ^t = \frac{1}{\sqrt {F\left( r \right)} }\left(
{{\begin{array}{*{20}c}
 i \hfill & 0 \hfill \\
 0 \hfill & { - i} \hfill \\
\end{array} }} \right),
\quad \gamma ^r = \sqrt {G\left( r \right)} \left(
{{\begin{array}{*{20}c}
 0 \hfill & {\sigma ^3} \hfill \\
 {\sigma ^3} \hfill & 0 \hfill \\
\end{array} }} \right),  \nn\\
&&\gamma ^\theta = \frac{1}{K\left( r \right)}\left(
{{\begin{array}{*{20}c}
 0 \hfill & {\sigma ^1} \hfill \\
 {\sigma ^1} \hfill & 0 \hfill \\
\end{array} }} \right),
\quad \gamma ^\phi = \frac{1}{H\left( r \right)}\left(
{{\begin{array}{*{20}c}
 0 \hfill & {\sigma ^2} \hfill \\
 {\sigma ^2} \hfill & 0 \hfill \\
\end{array} }} \right),
\eea
where $\sigma^i(i=1,2,3)$ is Pauli matrices. Being Dirac particles with spin $1/2$, the
wave function must be well described with both spin up ($ \uparrow
)$ and spin down ($\downarrow )$. Here we choose the wave functions
with spin up ($ \uparrow $) and  spin down ($\downarrow $) taking
the form as \bea \label{eq81} &&\psi _{\left( \uparrow \right)} =
\left( {{\begin{array}{*{20}c}
 A(t,r,\theta,\phi) \hfill \\
 0 \hfill \\
 B(t,r,\theta,\phi) \hfill \\
 0 \hfill \\
\end{array} }} \right)\exp \left( {\frac{i}{\hbar }I_ \uparrow(t,r,\theta,\phi) } \right),\nn\\
&&\psi _{\left( \downarrow \right)} = \left(
{{\begin{array}{*{20}c}
 0 \hfill \\
 C(t,r,\theta,\phi) \hfill \\
 0 \hfill \\
 D(t,r,\theta,\phi) \hfill \\
\end{array} }} \right)\exp \left( {\frac{i}{\hbar }I_ \downarrow(t,r,\theta,\phi) } \right).
\eea
Here, $I_\uparrow$ and $I_ \downarrow$ respectively denote the
action of Dirac particles with spin up ($ \uparrow $) and spin
down ($\downarrow $) tunnelling across black hole horizons. Inserting
the wave function with spin up ($\uparrow$) into Dirac equation
(\ref{eq6}) (the spin down case ($\downarrow $) proceeding in a manner fully
analogous to the spin up case$\uparrow$), then dividing by the exponential
term and multiplying by $\hbar$, the resulting equations to leading order
in $\hbar$ are
\begin{equation}
\label{eq8}
 - \left( {\frac{iA}{\sqrt {F\left( r \right)} }\partial _t I_ \uparrow +
B\sqrt {G\left( r \right)} \partial _r I_ \uparrow } \right) + mA =0,
\end{equation}
\begin{equation}
 \label{eq9} \left( {\frac{iB}{\sqrt {F\left( r \right)} }\partial _t
I_ \uparrow - A\sqrt {G\left( r \right)} \partial _r I_ \uparrow }
\right) + mB = 0,
\end{equation}
\begin{equation}
\label{eq10}
 - B\left( {\frac{1}{K\left( r \right)}\partial _\theta I_ \uparrow +
\frac{i}{H\left( r \right)}\partial _\phi I_ \uparrow } \right) = 0,
\end{equation}
\begin{equation}
\label{eq11}
 - A\left( {\frac{1}{K\left( r \right)}\partial _\theta I_ \uparrow +
\frac{i}{H\left( r \right)}\partial _\phi I_ \uparrow } \right) = 0.
\end{equation}

It is difficult to directly get the value of the action. From the above
equations, we can find the action can be separated. Considering
the symmetries of Kerr de Sitter black hole, we carry out
separation of variables as
\begin{equation}
\label{eq12} I_ \uparrow = - (\omega-j\Omega) t + W\left( r\right) + j\phi + \Theta \left( \theta \right),
\end{equation}

where $\omega $ is the energy of the emitted particles measured by
the observer at the infinity, and $j$ denotes the angular quantum
number about $\varphi$.
Inserting Eq.(\ref{eq12}) into Eqs.(\ref{eq8}),
(\ref{eq9}), (\ref{eq10}) and (\ref{eq11}) , we have
\begin{equation}
\label{eq13} \left( {\frac{iA}{\sqrt {F\left( r \right)} }\left(
{\omega - j\Omega } \right) - B\sqrt {G\left( r \right)} \partial_r
W\left( r \right)} \right) + mA = 0,
\end{equation}
\begin{equation}
\label{eq14}
 - \left( {\frac{iB}{\sqrt {F\left( r \right)} }\left( {\omega - j\Omega }
\right) + A\sqrt {G\left( r \right)} \partial_r W\left( r \right)}
\right) + mB = 0.
\end{equation}
\begin{equation}
\label{eq100}
 - B\left( \frac{1}{K\left( r \right)}\partial_\theta \Theta\left(\theta\right) +
\frac{i}{H\left( r \right)}j \right) = 0,
\end{equation}
\begin{equation}
\label{eq110}
 - A\left( \frac{1}{K\left( r \right)}\partial_\theta \Theta\left(\theta\right) +
\frac{i}{H\left( r \right)}j \right) = 0.
\end{equation}

From Eqs.(\ref{eq100}) and (\ref{eq110}), $\Theta\left(\theta\right)$ must be
a complex function, which means, in computing the imaginary part of the action
it yields a contribution. However further studying indicates that the contribution from $\Theta\left(\theta\right)$
is completely same for both the outgoing and ingoing solutions. So when dividing the outgoing
probability by the incoming probability, the contribution of $\Theta\left(\theta\right)$ to
the imaginary part of the action is cancelled out.
Although there are four equations, we are only interested
in the first two, for the imaginary part of the action can be produced here.
Now the tunnelling rate is directly related to the solutions of
Eqs.(\ref{eq13}) and (\ref{eq14}). When $m = 0$, the above equations
describe Hawking radiation for massless particles, and
Eqs.(\ref{eq13}) and (\ref{eq14}) decouple. While $m \ne 0$, they describe Hawking
radiation for massive particles, and Eqs.(\ref{eq13}) and (\ref{eq14}) couple. In fact,
whatever Eqs.(\ref{eq13}) and (\ref{eq14}) couple or not, it doesn't
affect the result we want. Eqs.(\ref{eq13}) and (\ref{eq14}) have a non-trivial solution for
A and B only if the determinant of the  coefficient matrix vanishes, which results
\begin{equation}
\label{eq15} W_\pm\left( r \right) = \pm \int {\sqrt {\frac{\left(
{\omega - j\Omega } \right)^2 +m^2F\left( r \right)}{F\left( r
\right)G\left( r \right)}} dr},
\end{equation}
where $+/- $ sign corresponds to the outgoing/ingoing wave of the particles. Now,
we focus on considering Hawking radiation at the EH of the black hole. At the EH, solving
Eq.(\ref{eq15}) yields
 \bea
 \label{eq16}
 W_{h\pm}\left( r \right) &=& \pm i\pi \frac{\omega - j\Omega _h
 }{\sqrt {{F}'\left( {r_h } \right){G}'\left( {r_h } \right)} } \nn\\
 &=& \pm i\pi \frac{\left( {r_h^2 + a^2} \right)\left( {\omega - j\Omega _h }
 \right)}{2\left( {r_h - 2r_h^3 l^{ - 2} - r_h a^2l^{ - 2} -
 M} \right)},
\eea

 and $\Omega _h =\Omega(r_h)= {a\Xi }/({r_h^2 + a^2})$ is the
 angular velocity at the event horizon(EH) of the black hole.
Then the tunnelling probability for Dirac particles with spin up ($\uparrow$) across the EH
takes the form as

\bea
\label{eq17} \Gamma &=& \frac{P(emission)}{P(absorption)} = \frac{\exp
\left( { - 2ImI_{\uparrow + }} \right)}{\exp \left( { - 2ImI_ {\uparrow- }} \right)}
=\frac{\exp{\left(-2Im W_{h+}\right)}}{\exp{\left(-2Im W_{h-} \right)}} \nn\\
&= &\exp \left( { - 2\pi \frac{\left( {r_h^2 + a^2} \right)\left(
{\omega - j\Omega _h } \right)}{r_h - 2r_h^3 l^{ - 2} - r_h
a^2l^{ - 2} - M}} \right).
\eea
This is the Boltzmann factor with Hawking temperature at the EH of Kerr de Sitter black hole taking
\begin{equation}
\label{eq18} T_h = \frac{r_h -
2r_h^3 l^{ - 2} - r_h a^2l^{ - 2} - M}{2\pi\left(r_h^2 + a^2\right)},
\end{equation}

which is fully in accordance with that obtained by other
methods \cite{r9}. When $\Lambda = 3l^{ -2}=0$, from Eq.(\ref{eq18})
Hawking temperature of Kerr black hole is recovered\cite{r12}. When $a=0$ and
$\Lambda = 3l^{ -2}=0$, the temperature reads as $1/8\pi M$, which is exactly
Hawking temperature of Schwarzschild black hole.

Beyond particles tunnelling from the EH, particles can also
tunnel from the CH. Different from the tunnelling behavior
across the EH, the emitted particle is found to tunnel
into the CH. Next, our focus is on computing the emission rate of Dirac particles
via tunnelling from the cosmological horizon(CH). At the CH, according to Eq.(\ref{eq15}),
the radial wave function $W_c\left(r\right)$ satisfies
\bea
 \label{eq185}
 W_{c\pm}\left( r \right) &=& \pm i\pi \frac{\omega - j\Omega _c
 }{\sqrt {{F}'\left( {r_c } \right){G}'\left( {r_c } \right)} } \nn\\
 &=& \pm i\pi \frac{\left( {r_c^2 + a^2} \right)\left( {\omega - j\Omega _c }
 \right)}{2\left( {r_c - 2r_c^3 l^{ - 2} - r_c a^2l^{ - 2} -
 M} \right)},
\eea
where $r_c$ determines the location of the CH, and $\Omega _c =\Omega(r_c)= {a\Xi }/({r_c^2 + a^2})$ is
the angular velocity at the CH of the black hole. Considering particles tunnelling into
the CH, the emission rate for Dirac particles with spin up ($\uparrow$) across the CH is
\bea
\label{eq186}
\Gamma &=&\frac{P(absorption)}{P(emission)}= \frac{\exp
\left( { - 2ImI_{\uparrow -} } \right)}{\exp \left( { - 2ImI_ {\uparrow +} } \right)}
 =\frac{\exp \left(  - 2ImW_{c-}\right)}{\exp\left(  - 2ImW_{c+}\right)} \nn\\
 &= &\exp \left( {  2\pi \frac{\left( {r_c^2 + a^2} \right)\left(
 {\omega - j\Omega _c } \right)}{r_c - 2r_c^3 l^{ - 2} - r_c
 a^2l^{ - 2} - M}} \right).
\eea
Here, to let Eq.(\ref{eq186}) taken the form as the Boltzmann factor (that is,
$\Gamma\propto\exp{\left(-\beta_c E\right)}$, where $\beta_c$ is the inverse temperature at the CH ),
two choices may appear to us. One choice is taking the negative inverse temperature $\beta_c$, and
the other considering $E=- \left(\omega - j\Omega _c \right)$. However the negative temperature
seems to be unphysical. So we only choose to change the sign of the positive "frequency", that is to say, when
measured an emitted particle taking a positive "frequency", its energy becomes negative in de Sitter spaces.
This is similar to the case that in de Sitter spaces the positive mass $m$ is measured as the negative energy
by $E=-m$. This new interpretation of the emission rate recently appears in Ref.\cite{r10}. As a result,
Hawking temperature at the CH of Kerr de Sitter black hole can be written as
\begin{equation}
\label{eq187} T_c = \frac{r_c -
\mbox{2}r_c^3 l^{ - 2} - r_c a^2l^{ - 2} - M}{2\pi\left(r_c^2 + a^2\right)}.
\end{equation}
This result is consistent with that in Ref.\cite{r9}. Obviously, Hawking temperatures of
Dirac particles tunnelling from the EH and the CH of Kerr de Sitter black hole are well
described via fermions tunnelling method. In our case, we only consider tunnelling particles
with spin up, but for spin down case, we can also adopt the same procedure to get the same result.

\section{Charged Dirac particles tunnelling from Kerr-Newman de Sitter black hole} \label{knds}

In this section, to further develop Kerner and Man's fermion tunnelling method, taking a charged rotating
(Kerr-Newman) black black hole in de Sitter space as an example, we study Hawking radiation of charged
Dirac particles tunnelling from the EH and the CH. The metric of Kerr-Newman de Sitter black hole \cite{r11}
can be given by Eqs.(\ref{eq1}) and (\ref{eq2}) by replacing $2Mr$ with $(2Mr - Q^2)$. Correspondingly, its
electromagnetic potential takes the following form as
\begin{equation}
\label{eq19}
A_\mu = A_t dt + A_\varphi
d\varphi = \frac{Qr}{r^2 + a^2\cos ^2\theta }dt - \frac{Qra\sin
^2\theta }{\left( {r^2 + a^2\cos ^2\theta } \right)\Xi }d\varphi .
\end{equation}

As mentioned in Sec.\ref{kds}, for simplicity, we suppose all the
observers are dragged with the rotating black hole, which means
carrying out the dragging coordinate transformation $\phi = \varphi
- \Omega t$, where
\begin{equation}
\label{eq2003}
\Omega = \frac{\left[ {\Delta _\theta \left( {r^2 + a^2} \right) -
\widetilde{\Delta} } \right]a\Xi }{\Delta _\theta \left( {r^2 + a^2} \right)^2 -
\widetilde{\Delta} a^2\sin ^2\theta },
\end{equation}
is the angular velocity of the black
hole. After performing this transformation, the new metric takes
the same form as Eq.(\ref{eq4}) only replacing $\Delta$ with
$\widetilde{\Delta}=\left( {r^2 + a^2} \right)\left( {1- r^2l^{ - 2}} \right) - 2Mr + Q^2$,
and the electromagnetic potential correspondingly changes as
\begin{equation}
\label{eq20} \mathcal{A}_\mu = \mathcal{A}_t dt + \mathcal{A}_\phi d\phi = \frac{\Delta _\theta
Qr\left( {r^2 + a^2} \right)}{\Delta _\theta \left( {r^2 + a^2}
\right)^2 - \widetilde{\Delta }a^2\sin ^2\theta }dt - \frac{Qra\sin
^2\theta }{\left( {r^2 + a^2\cos ^2\theta } \right)\Xi }d\phi .
\end{equation}

When dealing with charged Dirac particles tunnelling from black holes, we
must first introduce its motion equation. In the curved spaces, charged Dirac particles
satisfy Dirac equation as
\begin{equation}
\label{eq21} i\gamma ^\mu \left( {\partial _\mu + \Omega _\mu +
\frac{i}{\hbar }e\mathcal{A}_\mu } \right)\psi + \frac{m}{\hbar }\psi = 0,
\end{equation}
where $e$ is the electric charge of the tunnelling particle, and $\mathcal{A}_\mu $
represents the electromagnetic potential of the black hole. Considering the metrics in Sec.\ref{kds} and \ref{knds} sharing
the similar forms, here we choose the matrices $\gamma ^\mu $ taking the form as Eq.(\ref{eq7}), and the
wave functions with spin up ($ \uparrow $) and spin down ($\downarrow $) as Eq.(\ref{eq81}).
For simplicity in our discussion, we only consider the tunnelling particles taking spin up ($ \uparrow $). Substituting
the wave function with spin up ($ \uparrow $) into Dirac equation (\ref{eq21}), we have
\begin{equation}
\label{eq22}
 - \left( {\frac{iA}{\sqrt {F\left( r \right)} }\left( {\partial _t I_
\uparrow + e\mathcal{A}_t } \right) + B\sqrt {G\left( r \right)} \partial _r
I_ \uparrow } \right) + mA = 0,
\end{equation}
\begin{equation}
\label{eq23} \left( {\frac{iB}{\sqrt {F\left( r \right)} }\left(
{\partial _t I_ \uparrow + e\mathcal{A}_t } \right) - A\sqrt {G\left( r
\right)} \partial _r I_ \uparrow } \right) + mB = 0,
\end{equation}
\begin{equation}
\label{eq24}
 - B\left( {\frac{1}{K\left( r \right)}\partial _\theta I_ \uparrow +
\frac{i}{H\left( r \right)}\left( {\partial _\phi I_ \uparrow +
e\mathcal{A}_\phi } \right)} \right) = 0,
\end{equation}
\begin{equation}
\label{eq25}
 - A\left( {\frac{1}{K\left( r \right)}\partial _\theta I_ \uparrow +
\frac{i}{H\left( r \right)}\left( {\partial _\phi I_ \uparrow +
e\mathcal{A}_\phi } \right)} \right) = 0.
\end{equation}

Due to the presence of the electric charge for the emitted particles, the above equations
are different from those in Sec.(\ref{kds}), but the action can also be
separated. Our interest is still the first two. Being taken the same symmetries
between Kerr de Sitter black hole and Kerr-Newman de Sitter black hole, we carry out
separation of variables as Eq.(\ref{eq12}) to Eqs.(\ref{eq22}) and (\ref{eq23}), which yields
\begin{equation}
\label{eq26} \left( {\frac{iA}{\sqrt {F\left( r \right)} }\left(
{\omega - e\mathcal{A}_t - j\Omega } \right) - B\sqrt {G\left( r \right)}
\partial _r W\left( r \right)} \right) + mA = 0,
\end{equation}
\begin{equation}
\label{eq27}
 - \left( {\frac{iB}{\sqrt {F\left( r \right)} }\left( {\omega - e\mathcal{A}_t -
j\Omega } \right) + A\sqrt {G\left( r \right)} \partial _r W\left( r
\right)} \right) + mB = 0.
\end{equation}

As in Sec.(\ref{kds}), here we do not consider the contribution of the angular part to the
imaginary part of the action. To demand Eqs.(\ref{eq26}) and (\ref{eq27}) solvable,
the determinant of the coefficient matrix must be vanished, that is
\begin{equation}
\label{eq28} W_\pm \left( r \right) = \pm \int {\sqrt {\frac{\left(
{\omega - e\mathcal{A}_t - j\Omega } \right)^2 + m^2F\left( r
\right)}{F\left( r \right)G\left( r \right)}} dr},
\end{equation}
where $+/- $ correspond to the outgoing/ingoing solutions. At the
EH of the black hole, solving Eq.(\ref{eq28}), we have
\bea
\label{eq128} W_{h\pm} \left( r \right)&=&\pm i\pi\frac{\omega -
e\mathcal{A}_t(r_h)-j\Omega _h }{\sqrt{F'\left(r_h\right)G'\left(r_h\right)}} \nn\\
&=& \pm i\pi \frac{\left(
{r_h^2 + a^2} \right)\left( {\omega - e\mathcal{A}_t(r_h) -
j\Omega _h } \right)}{2\left( {r_h - 2r_h^3 l^{ - 2} - r_h a^2l^{
- 2} - M} \right)},
\eea
where $\Omega _h= \Omega(r_h)= {a\Xi
}/\left({r_h^2 + a^2}\right)$ is the angular velocity at the EH of
Kerr-Newman de Sitter black hole. So the tunnelling probability of Dirac particles across
the event horizon(EH) is
\bea
\label{eq30} \Gamma &=& \frac{P(emission)}{P(absorption)} = \frac{\exp
\left( { - 2ImI_{\uparrow + }} \right)}{\exp \left( { - 2ImI_ {\uparrow- }} \right)}
=\frac{\exp{\left(-2Im W_{h+}\right)}}{\exp{\left(-2Im W_{h-} \right)}} \nn\\
&= & \exp \left( { - 2\pi\frac{\left( {r_h^2 + a^2} \right)\left( {\omega - e\mathcal{A}_t(r_h) -
j\Omega _h } \right)}{r_h - 2r_h^3 l^{ - 2} - r_h a^2l^{ - 2}
- M}} \right).
\eea
And Hawking temperature at the EH of Kerr-Newman de Sitter black hole
is recovered as
\begin{equation}
\label{eq31} T_h = \frac{1}{2\pi }\frac{r_h -
2r_h^3 l^{ - 2} - r_h a^2l^{ - 2} - M}{r_h^2 + a^2}.
\end{equation}
This result is consistent with that in Ref.\cite{r11}.
When $\Lambda = 3l^{ -2}=0$, from Eq.(\ref{eq31})
Hawking temperature of Kerr-Newman black hole is recovered\cite{r12,r13}. When $a=0$, $Q=0$ and
$\Lambda = 3l^{ -2}=0$, Kerr-Newman de Sitter black hole is reduced to Schwarzschild black hole, and the temperature
is recovered as $1/8\pi M$.

As in Sec.(\ref{kds}), Hawking radiation can also be produced at the cosmological horizon(CH).
At the CH, outgoing particles can fall out of the horizon along classically permitted trajectories, and
incoming particles tunnel from it to form Hawking radiation.
This tunnelling behavior is different from that at the EH,
where incoming particles fall behind the horizon along classically permitted trajectories, and Hawking
radiation is produced by outgoing particles tunnelling from the horizon. As a result, taking the same
procedure as Sec.(\ref{kds}), Hawking temperature at the CH of Kerr-Newman de Sitter black hole
is
\begin{equation}
\label{eq32} T_c = \frac{1}{2\pi }\frac{r_c -
2r_c^3 l^{ - 2} - r_c a^2l^{ - 2} - M}{r_c^2 + a^2}.
\end{equation}
This result takes the same form as that in Ref.\cite{r11}. So Hawking temperature of Kerr-Newman
de Sitter black hole can also be recovered by charged Dirac particles tunnelling from the EH and the CH.
As before, here we only consider the tunnelling Dirac particles with spin up, for the spin down case, the
same result can be recovered.

\section{Conclusion and Discussion}\label{SSDB}
In this paper, considering Dirac particles tunnelling from the event horizon(EH) and the cosmological
horizon(CH), we have studied Hawking radiation from rotating black holes in de Sitter spaces, which includes that from Kerr
de Sitter black hole and Kerr-Newman de Sitter black hole. As a result, Hawking temperatures at the EH and the CH of the
two rotating black holes are successfully recovered. These results promote Kerner and Man's fermion tunnelling method to the cases
of Dirac particles tunnelling from the EH and the CH of rotating black holes in de Sitter spaces, and once again prove its
validity and universality.

Different from the tunnelling behavior at the EH, at the CH
outgoing particles can fall out of the horizon
along classically permitted trajectories, and incoming particles tunnel from it to form Hawking radiation.
So the tunnelling rate should be expressed by dividing the incoming probability by the outgoing probability
as Eq.(\ref{eq186}). On the other hand, when doing de Sitter radiation, to set the tunnelling rate taken as the
Boltzmann factor, we consider $E=- \left(\omega - j\Omega  \right)$, which means an emitted particle
taking a positive "frequency", its energy becomes negative in de Sitter spaces. This new interpretation
of the emission rate can be recently seen in Ref.\cite{r10}.

When carrying on separation of variables for the action, we only concern ourselves with positive frequency contributions.
And when considering Dirac particles tunnelling from the EH and the CH of black holes, we only refer to those with spin up.
In fact, these two hypothesis would not stop fermion tunnelling method from losing its generality.

\section*{Acknowledgments}
This work is supported by National Natural Science Foundation of
China (Grant Nos.10675051, 10705008).

$Note~added-$
As this work was being complete, the papers \cite{M9}, \cite{M10} and \cite{M11} appeared,
respectively discussing fermions tunnelling from Kerr black hole, Kerr-Newman black hole and dynamical horizon.


\begin{thebibliography}{99}

\bibitem{r1}

S. W. Hawking, Nature \textbf{30} (1974) 248.

\bibitem{r2}

P. Kraus and F. Wilczek, Nucl. Phys. B \textbf{437} (1995) 231;
\textbf{433} (1995) 403.

\bibitem{r3}

M. K. Parikh, and F. Wilczek, Phys. Rev. Lett. \textbf{85} (2000) 5042.

\bibitem{r4}
M. Angheben, M. Nadalini, L. Vanzo and S. Zerbini, JHEP
\textbf{05} (2005) 014.

\bibitem{r5}
K. Srinivasan and T. Padmanabhan, Phys. Rev. D \textbf{60} (1999)
024007;

S. Shankaranarayanan, K. Srinivasan, and T. Padmanabhan, Mod.
Phys. Lett. A \textbf{16} (2001) 571;

S. Shankaranarayanan, T. Padmanabhan, and K. Srinivasan, Class.
Quant. Grav. 19, 2671 (2002);

\bibitem{r6}
S. Hemming and E. Keski-Vakkuri, Phys. Rev. D \textbf{64} (2001) 044006;

A. J. M. Medved, Phys. Rev. D \textbf{66} (2002) 124009;  Class. Quant.
Grav. \textbf{19} (2002) 589;

E. C. Vagenas, Phys. Lett. B \textbf{503} (2001) 399; Phys. Lett. B \textbf{559} (2003) 65; Mod. Phys.
Lett. A \textbf{17} (2002) 609;

A. J. M. Medved and E. C. Vagenas, Mod. Phys. Lett. A \textbf{20} (2005) 1723; Mod. Phys. Lett. A \textbf{20} (2005) 2449;

M. Arzano, A. J. M. Medved and E. C. Vagenas, JHEP \textbf{0509} (2005) 037;

J. Y. Zhang and Z. Zhao, Phys. Lett. B \textbf{618} (2005) 14; JHEP \textbf{0505} (2005) 055;

Q. Q. Jiang and S. Q. Wu, Phys. Lett. B \textbf{635} (2006) 151;

S. Q. Wu and Q. Q. Jiang, JHEP, \textbf{0603} (2006) 079;

Y. P. Hu, J. J. Zhang and Z. Zhao, Mod. Phys. Lett. A \textbf{21} (2006) 2143;

S. P.Kim, JHEP \textbf{0711} (2007) 048;

S. Sarkar and D. Kothawala, Phys. Lett. B \textbf{659} (2008) 683;

T. Pilling, Phys. Lett. B \textbf{660} (2008) 402;

M. H. Ali, Class. Quant. Grav.  \textbf{24} (2007) 5849;

R. Kerner and R. B. Mann, Phys. Rev. D \textbf{75} (2007) 084022;
Phys. Rev. D \textbf{73} (2006) 104010;

C. Z. Liu, J. Y. Zhang and Z. Zhao, Phys. Lett. B \textbf{639}
(2006) 670;

W. B. Liu, Phys. Lett. B \textbf{634} (2006) 541;

P. Mitra, Phys. Lett. B \textbf{648} (2007) 240;

D. Y. Chen and S. Z. Yang, Ger. Rel. Grav. \textbf{39} (2007)
1503;

R. Banerjee and B. R. Majhi, Phys. Lett. B \textbf{662} (2008) 62;

R. Banerjee and S. Kulkarni, Phys. Lett. B \textbf{659} (2008) 827.

\bibitem{r7}

R. Kerner and R. B. Mann, arXiv:0710.0612 [hep-th].

\bibitem{M12}

R. Li and J. R. Ren, Phys. Lett. B \textbf{661} (2008) 370;

D. Y. Chen, Q. Q. Jiang, S. Z. Yang and X. T. Zu,  arXiv:0803.3248 [hep-th].

\bibitem{r14}

A.G. Reiss, et al., Astron. J. \textbf{116} (1998) 1009;

S. Perlmutter, et al., Astrophys. J. \textbf{517} (1999) 565;

J. P. Ostriker, P. J. Steinhardt, Nature \textbf{377} (1995) 600.

\bibitem{r15}

A. Strominger, JHEP \textbf{0110} (2001) 034;

D. Klemm, Nucl. Phys. B \textbf{625} (2002) 295.

\bibitem{r8}

B. Carter, Commun. Math. Phys. \textbf{10} (1968) 280.

\bibitem{r9}

S. Wang, S. Q. Wu, F. Xie and L. Dan, Chin. Phys. Lett. \textbf{10} (2006) 1096;

Q. Q. Jiang, Class. Quant. Grav. \textbf{24} (2007) 4391.

\bibitem{r10}

Y. Sekiwa, arXiv:0802.3266 [hep-th.].

\bibitem{r11}

M. H. Dehgham and H. KhajehAzad, hep-th/0209203.

\bibitem{r12}
Q. Q. Jiang, S. Q. Wu and X. Cai, Phys. Rev. D  \textbf{73} (2006) 064003; Erratum-ibid. \textbf{73} (2006) 069902.

\bibitem{r13}
J. Y. Zhang and Z. Zhao, Phys. Lett. B \textbf{638} (2006) 110.

\bibitem{M9}
R. Li and J.R. Ren, arXiv:0803.1410 [hep-th].

\bibitem{M10}
R. Kerner and R.B. Mann, arXiv:0803.2246 [hep-th].

\bibitem{M11}
R. D. Criscienzo and L. Vanzo, arXiv:0803.0435 [hep-th].

\end{thebibliography}
\end{document}